\documentstyle[12pt]{article}
\begin{document}
\baselineskip=24pt
\centerline {\bf Symmetrized DMRG Method for Excited States}
\centerline {\bf of Hubbard Models}

\centerline { S. Ramasesha$^{1,3}$, Swapan K. Pati$^{1}$, H. R. Krishnamurthy
 $^{2,3}$}
\centerline {and}
\centerline  {Z. Shuai$^{4}$ and J. L. Br{\' e}das$^{4}$} 
\centerline {\it ${^1}$Solid State and Structural Chemistry Unit}
\centerline {\it ${^2}$Department of Physics} 
\centerline {\it Indian Institute Of Science, Bangalore 560012 , India.}
\centerline {\it ${^3}$Jawaharlal Nehru Center for Advanced Scientific
Research}
\centerline {\it Jakkur Campus, Bangalore 560064, India.}
\centerline {\it ${^4}$Centre de Recherche en Electronique et
Photonique Mol{\' e}culaires}
\centerline {\it Service de Chimie des Mat{\' e}riaux Nouveaux}
\centerline {\it Universit{\' e} de Mons-Hainaut, Place du Parc 20, 7000 Mons, Belgium.}
\begin{center}
{\bf Abstract}
\end{center}
We extend the density matrix renormalization group
method to exploit Parity, $C_2$ (rotation by $\pi$) and
electron-hole symmtries of model Hamiltonians. We demonstrate the 
power of this method by obtaining  the lowest energy states 
in all the eight symmetry subspaces of Hubbard chains with upto 
50 sites. The ground-state energy, optical gap and spin gap of regular 
$U=4t$ and $U=6t$ Hubbard chains agree very well with exact results.
This development extends the scope of the DMRG method and allows future
applications to study of optical properties of low-dimensional conjugated
polymeric systems. 
\vskip 0.2cm
\noindent
PACS Numbers: 71.10.+x, 71.28.+d  and 78.20.-e

\pagebreak
The Density Matrix Renormalization Group (DMRG) method [1, 2] has 
proved to be an
accurate technique for obtaining a few low-lying states of interacting model
Hamiltonians with short-range interactions in low-dimensions[3-5]. This
technique would find wide applications in the electronic
structure studies of conjugated polymers and donor-acceptor systems,
if we can obtain excited eigenstates in different symmetry subspaces.
However, the DMRG techniques reported in the literature for fermions 
can  exploit only the 
conservation of the z-component of the total Spin ($S^{z}_{tot}$) and 
total number of particles($N_{tot}$) in the system[6-8]. Using these 
symmetries one can obtain a few (about ten)
low-lying states in different $M_s$ and $N$ sectors[2].  This is  due 
to the limitation on the number of 
eigenstates that can be targetted effectively because of the large
matrices that one encounters in the DMRG algorithm.
Additionally, the Hamiltonians being spin conserving, the states with higher
total spin  repeat in many $M_{s}$ sectors. For example, triplet states 
will be present in the  $M_{s}=0,+1$ and $-1$ sectors. 
Therefore, obtaining the second singlet state usually
requires targetting the third state in the $M_{s}=0$ sector and 
eliminating the $M_s = 0$ intruder triplet state.  
Besides, with increasing system size, the number of intruders also 
increases. Thus the inclusion of 
just these symmetries is not sufficient to make detailed 
comparisons between theory
and experiments that probe excited states.

In system with spatial symmetries, the states important from 
the point of view of optical (one-photon) spectroscopies  always 
lie in a spatial symmetry subspace different from the subspace in
which the  ground state is found. For example, in Hubbard chains, 
the ground state is in the A subspace while the dipole allowed excited 
states are in the B subspace [9].  

Hubbard chains at half-filling also possess electron-hole 
symmetry. This allows labelling of the states as either  
$'+'$, corresponding to {\it covalent} subspace, or $'-'$, corresponding to
{\it ionic} subspace.  The dipole allowed excited states are 
found in the $'-'$ subspace while the ground state is in the $'+'$ subspace. 
The number of higher energy states in the $'+'$ subspace intruding 
below the lowest $'-'$ subspace state increases
as the strength of electron correlation $U/t$ increases [10].
Hence, if this
symmetry is not incorporated, many $'+'$ space states intrude below the lowest
$'-'$ space state rendering the optically allowed excited state difficult 
to access computationally . 

Obtaining the Hamiltonian in fully block-diagonal form, with respect
to the above symmetries, is also
necessary, if we wish to compute the dynamic nonlinear optic
coefficients using correction vectors[11]. The correction vectors belong to
symmetry determined subspaces and are obtained  by solving a 
set of inhomogeneous linear algebraic equations which are defined 
by the Hamiltonian matrix. For example, the correction vector $\phi^{(1)}_{i}
(\omega)$, required for computing first order nonlinearities, is
defined by 
  $$ (H -E_{G} + \omega)\phi^{(1)}_{i}(\omega) = \hat{\mu}_{i}|G\rangle \eqno(1)$$
\noindent where $\hat{\mu}_{i}$ is the dipole displacement operator, $E_{G}$ and
$|G\rangle$ are the
ground state energy and eigenvector, and $\omega$ is the excitation frequency.
The vector $\phi^{(1)}_{i}(\omega)$
lies in the $^{1}B^{+}$ space, if the Hamiltonian has $C_{2}$, electron-hole
and spin  symmetries. If the Hamiltonian matrix is not block
diagonal then for small values of $\omega$ the system is near singular
making it almost impossible to obtain the solution, $\phi^{(1)}_{i}(\omega)$. 

In this communication, we briefly outline a new, {\it symmetrized DMRG}
 procedure which exploits fully the above symmetries in the context
of DMRG. As an example of its utility, we present
 some excitation gaps for long Hubbard chains and compare
them with the Bethe-{\it ansatz} solutions where possible.

In the implementation of the symmetrized DMRG scheme, we use the group
theoretic projection operators for projecting the direct product functions
onto different symmetry subspaces. The symmetries we have incorporated
are the electron-hole symmetry, $\hat J$,  end-to-end 
interchange symmetry of chains, $\hat {C}_{2}$ and parity, $\hat P$. 
The latter two symmetries have been employed within the DMRG scheme for
spin chains.
While the full spin symmetry 
classification can be incorporated into the scheme using
Clebsch-Gordon coefficients, for the present we have  
only included parity which bifurcates the space of spin eigenvectors 
into  even (e) and  odd (o) spin spaces.

The electron-hole symmetry operator interchanges, with a phase, the creation
and annihilation operators at a site,
$$a^{\dagger}_{i}=(-1)^{i} b_{i} \eqno(2) $$
\noindent Thus, the Fock space of a single site i, under electron-hole 
symmetry transforms as $\hat {J}_{i}|0> = |\uparrow \downarrow>$,
$\hat {J}_{i}|\uparrow>
= (-1)^{i}|\uparrow>$, $\hat {J}_{i}|\downarrow> = (-1)^{i}|\downarrow>$,
$\hat {J}_{i}|\uparrow \downarrow>=(-1)|0>$.
 The full electron-hole symmetry operator, $\hat{J}$, is given by the 
direct product of the single site operators.
$${\hat J}=\prod_{i} {\hat J}_{i} \eqno(3) $$
The parity operator, at a site i, ${\hat P}_{i}$, flips the electron spin
at a site. Thus, ${\hat P}_{i}|0> = |0>$, ${\hat P}_{i}|\downarrow> = 
|\uparrow>$, ${\hat P}_{i}|\uparrow> = |\downarrow>$, ${\hat P}_{i}|\uparrow 
\downarrow > = (-1) |\uparrow \downarrow >$.
The full parity operator for the system is also given by the direct product
of the single site parity operators.
$${\hat P}=\prod_{i} {\hat P}_{i} \eqno(4) $$
The $\hat {C}_{2}$ symmetry interchanges the states of the left and right
halves of the system with a phase factor. Thus, $\hat {C}_{2}$ operating 
on the direct product state
 $|\mu,\sigma,\sigma^{\prime},\mu^{\prime}>$  gives,
$$\hat {C}_{2}|\mu,\sigma,\sigma^{\prime},\mu^{\prime}>= (-1)^{\gamma}
|\mu^{\prime},\sigma^{\prime},\sigma,\mu> ~; ~\gamma = ( n_{\mu}
+n_{\sigma} )(n_{\mu^{\prime}}+n_{\sigma^{\prime}} ), \eqno(5) $$
\noindent where, $\mu (\mu^{\prime})$ refers to the 
$\mu^{th} (\mu^{{\prime}{th}})$  eigenvector of the left (right) density
matrix while $\sigma (\sigma^{\prime})$ refers to the Fock space state 
of the new left (right) site as in the standard DMRG procedure,
 and $n_{\mu}, ~n_{\mu^{\prime}}, ~n_{\sigma}$ ~and~ 
$n_{\sigma^{\prime}}$ are the occupancies in the states $~|\mu >,
~|\mu^{\prime} >, ~|\sigma >$ and $~|\sigma^{\prime} >$ respectively. 

The operators $\hat{C}_{2}$, $\hat {J}$ and $\hat {P}$ commute 
amongst each other and the group 
formed by identity and these symmetries consists of eight elements 
with the remaining four elements, resulting from 
the closure condition. This group being abelian has eight irreducible 
representations which are labelled  $^{e}A^{\bf +}$, $^{e}A^{\bf -}$, 
$^{o}A^{\bf +}$, $^{o}A^{\bf -}$, $^{e}B^{\bf +}$, $^{e}B^{\bf -}$, 
$^{o}B^{\bf +}$, $^{o}B^{\bf -}$. 

The projection operator for a  given irreducible representation, $\Gamma$, 
is given by 
 $${\bf {\hat P}}_{\Gamma} = {1\over h} \sum_{\hat R} {\chi}_{_{\Gamma}}
({\hat R}) {\hat R} \eqno(4) $$  
\noindent where ${\hat R }$'s are the symmetry operations,
$\chi_{_{\Gamma}}{(\hat {R})}$ is the character of $\hat {R}$ in 
$\Gamma$ and $h$ is the order 
of the group. The construction of the {\it symmetry adapted} direct product 
states consists in
sequentially operating on each of the direct product states by the 
projection operator. The
linear dependencies of the symmetry adapted combinations that result
are eliminated by carrying out a Gram-Schmidt orthonormalization.

The computational procedure involves obtaining the matrix representation
of the symmetry operators of the $(2n+2)$ site chain in the 
direct product basis. The matrix 
representation of both $\hat J$ and $\hat P$ for the 
new sites in the Fock space is known
from  their definitions. Similarly, the matrix representation of the 
operators $\hat J$ and $\hat P$ for the
left (right) part of the system at the first iteration are also known
in the basis of the corresponding Fock space 
states. These are then transformed to the density
matrix eigenvectors basis. The matrix representation of the symmetry 
operators of the full system in the 
direct product space are obtained as the direct product of the corresponding
matrices,
  $$<\mu,\sigma,\sigma^{\prime},\mu^{\prime} | {\hat R}_{2n+2} | \nu,\tau,
  \tau^{\prime},
  \nu^{\prime}>=<\mu| {\hat R}_{n} |\nu> <\sigma | {\hat R}_{1}
  |\tau > <\sigma^{\prime}
  | {\hat R}_{1}|\tau^{\prime} > < \mu^{\prime}| {\hat R}_{n} 
  | \nu^{\prime} > \eqno(5) $$
At the next iteration, we require the matrix representation of the symmetry
operators in the basis of the eigenvectors of the new density
matrix. This is achieved by obtaining the matrix ${\bf R}$ as a direct product of
${\bf R}_{n}$ and ${\bf R}_{1}$ given by
  $$<\mu,\sigma |{\hat R}_{n+1}|\nu,\tau > = <\mu | {\hat R}_{n}|\nu> <\sigma |
  {\hat R}_{1}|\tau > \eqno(6) $$
\noindent This matrix is renormalized by the transformation
  $${ \tilde {\bf R} }_{n+1} = {\bf O}^{\dagger} {\bf R}_{n+1} {\bf O}, \eqno(7) $$
\noindent where ${\bf O}$ is the matrix whose columns are the chosen 
eigenvectors of the density matrix.
  
The matrix representation of $\hat {C}_{2}$ is quite straightforward in that each
state $|\mu,\sigma,\sigma^{\prime},\mu^{\prime}>$ is mapped into a state 
$|\mu^{\prime},\sigma^{\prime},\sigma,\mu>$ with a 
phase factor. Thus, each row in the matrix of $\hat{C}_{2}$ contains only one 
element and the entire matrix can be represented by a correspondence vector.
The  matrices ${\bf J}$ and ${\bf P}$ are also rather sparse and can be stored in sparse form to
avoid doing arithmetic with zeros. This aspect of the computation turns out to 
be crucial for the
implementation of the scheme since the dimensionality of the direct product 
space is usually very large ( about $10^{5}$). 

The coefficients of the direct product functions in the symmetrized basis
forms a matrix ${\bf S}$. The Hamiltonian matrix in the direct
product basis can be transformed to the Hamiltonian in the symmetrized
basis by the transformation
     
    $$ {\tilde {\bf H}}_{2n+2}= {\bf S}^{\dagger} {\bf H}_{2n+2} {\bf S} \eqno(8) $$

The size of the symmetrized Hamiltonian matrix is of the order 
$\simeq 2000 \times 2000$ for a cut-off of $m=100$.
\noindent The low-lying eigenstates of this matrix can be obtained by Davidson's 
algorithm.\\

The symmetry adaptation scheme described above has been implemented both
within the infinite chain DMRG algorithm and within the finite system 
algorithm. In the latter, we incorporate the $C_2$ symmetry only when the
left and right parts of the system are identical, {\it i.e.} at the end 
of each finite system iteration. Without iterating over 
the density matrices of the fragments ( i.e. within the infinite system 
algorithm), we find that the energy difference between a chain of length $N$
with $N+1$ and $N-1$ electrons is equal to the Hubbard parameter $U$ to 
an accuracy of $\approx 10^{-3}$. After three iterations of the 
finite system algorithm, the accuracy improves to $\approx 10^{-5}$, 
for a $U$ value of $4t$.

As an example of the use of the above technique, we present results of
DMRG calculations for uniform Hubbard chains 
at half-filling, for $U/t$ of 4.0 and 6.0, with chain lengths of upto 
50 sites. We have obtained the lowest energy states
in all the eight subspaces ( of ${\hat C}_2$, ${\hat J}$ and ${\hat P}$)
, keeping 70 to 150 eigenvectors of the 
density matrix. The extrapolated ground state energy per site 
of the infinite chain for both values of $U/t$
agrees with the exact values of $0.5737331t$ and $0.4200125t$ respectively
to 5 decimal places. 

The energy of the one-photon transition from the ground state
(lying in the $^{e}A^{+}$ subspace) to the lowest energy state in the 
$^{e}B^{-}$ subspace defines the optical gap of the chain. In fig.1 we 
show a plot of optical gap {\it vs.} inverse chain length for 
$U/t=4.0$ and
$6.0$ for different values of the cut-off m. The optical gaps for these
values of $U/t$, from the
Bethe {\it ansatz} solution for the infinite chain, are $1.2867t$ and
$2.8926t$ respectively [12]. The corresponding extrapolated values from
a polynomial fit in powers of $1/n$  from DMRG are $1.278t$ and $2.895t$, 
obtained with a cut-off of m=150 for $U/t=4.0 $
and m=100 for $U/t=6.0$. We also find that the DMRG optical
gap tends to saturate at shorter chain lengths as we decrease m. The fit
of the optical gap to $1/n$ is a reasonably good straight line although
the best fit is to a polynomial in this variable.

In fig.2 we plot the spin gap which we define as the energy gap
between the lowest triplet state and the ground state singlet as a function of
$1/n$. The Bethe {\it ansatz} solution yields a vanishing spin gap 
in the thermodynamic limit of the uniform Hubbard chain. Polynomial fits
to our DMRG data are consistent with this.

In fig.3 we present the energy gaps, relative to the ground state, obtained
by solving for the lowest energy state in each of the eight subspaces, for
Hubbard chains with 40 to 50 sites, for the two values of $U/t$. 
The excitations clearly break up into two bands. The lower energy 
excitations correspond to states of different symmetry 
in the {\it covalent} subspace ( i.e. predominantly spin excitations) 
, while the higher energy excitations correspond 
to the {\it ionic} subspace ( i.e. charge excitations) . 
With increasing $U/t$ the states in the {\it 
covalent} subspace are less dispersed and so are the states in 
the {\it ionic} subspace. However, with increase in $U/t$ the gap 
between the two bands increases. This feature is in agreement with
the basic physics of the Hubbard models.

To summarize, we have presented a symmetrized DMRG scheme for obtaining
excitation gaps in different symmetry subspaces of a large model system. 
We have illustrated our scheme with applications to a uniform Hubbard
model, wherein the known excitation gaps are very well reproduced. 
We expect that this scheme will prove invaluable in modelling the
electronic state properties of polymers.

\noindent Acknowledgement : We thank SERC ( IISc, India ) and JNCASR ( India )
for computational
facilities and Dr. Biswadeb Dutta for system help. The work in Mons is
supported by Belgian Prime Minister Science Policy Office, Program Pole
d'Attraction Interuniversitaire. Z. S. thanks Prof. Wu-Pei Su for helpful
discussions.

\pagebreak

\noindent{\bf References} \\

\begin{enumerate}
\item S. R. White, Phys. Rev. Lett. {\bf 69}, 2863 (1992) ; Phys. Rev. B
{\bf 48}, 10345 (1993).
\item R. M. Noack and S. R. White, Phys. Rev. B {\bf 47}, 9243 (1993)
\item S. R. White and D. Huse, Phys. Rev B {\bf 48}, 3844 (1993)
\item Erik S. Sorenson and Ian Affleck, Phys. Rev. B {\bf 49}, 15771 (1994)
\item R. Chitra, Swapan K Pati, H. R. Krishnamurthy, Diptiman Sen and
S. Ramasesha , Phys. Rev. B {\bf 52} 5681 (1995) 
\item Hanbin Pang, Shoudan Liang and James F. Annett, Phys. Rev. Lett., 
{\bf 71}, 4377 (1993).
\item Hanbin Pang and Shoudan Liang, Phys. Rev. B {\bf 51}, 10287 (1995).
\item S. R. White, R. M. Noack and D. J. Scalapino, Phys. Rev. Lett., 
{\bf 73}, 882 (1994).
\item Z. G. Soos and S. Ramasesha, Phys. Rev. B, {\bf 29}, 5410 (1984).
\item Z. G. Soos, S. Ramasesha and D. S. Galvao, Phys. Rev. Lett., {\bf 71}
, 1609 (1993).
\item Z. G. Soos and S. Ramasesha, J. Chem. Phys., {\bf 90}, 1067 (1989)
; S. Ramasesha and Z. G. Soos, Chem. Phys. Lett., {\bf 153}, 171 (1988).
\item E. H. Lieb and F. Y. Wu, Phys. Rev. Lett., {\bf 20}, 1445 (1968);
A. A. Ovchhinikov, Sov. Phys - JETP, {\bf 30}, 1160 (1970).
\end{enumerate}

\pagebreak
\begin{center}
{\bf Figure Captions:}
\end{center}

\noindent {\bf Fig.1 :} \\
Optical Gap as a function of inverse chain length for Hubbard chains 
with  $U=4.0t$ and $U=6.0t$. $m$ corresponds to the number of density
matrix eigenvectors retained in the DMRG procedure. Arrows indicate the
model exact gaps for infinite chains. \\
\noindent {\bf Fig.2 :} \\
Spin Gap (defined in the text) as a function of $1/n$ for Hubbard chains
with $U=4.0t$ and $U=6.0t$. $m$ corresponds to the DMRG cut-off. Model exact
spin gaps vanish for infinite chains. \\
\noindent {\bf Fig.3 :} \\
Energy gaps ( measured from the ground state) of the lowest state in each
subspace for chain length varying from  40 to 50, for two different values
of $U/t$. The level ordering is $E_{^{e}A^{\bf +}}$ < $E_{^{o}A^{\bf +}}$ < 
$E_{^{o}B^{\bf +}}$ < $E_{^{e}B^{\bf +}}$ < $E_{^{e}B^{\bf -}}$ < 
$E_{^{e}A^{\bf -}}$ < $E_{^{o}B^{\bf -}}$ < $E_{^{o}A^{\bf -}}$ .
\end{document}